\documentclass[epjST]{svjour}
\usepackage{graphicx}
\usepackage{amsmath}
\usepackage{amsfonts}
\usepackage{setspace}
\begin{document}

\title{Characteristics of in-out intermittency in delay-coupled FitzHugh-Nagumo oscillators}
\author{Arindam Saha\inst{1}\fnmsep\thanks{\email{arindam.saha@uni-oldenburg.de}} \and Ulrike Feudel\inst{1}}
\institute{Theoretical Physics/Complex Systems, ICBM, University of Oldenburg, 26129 Oldenburg, Germany}
\abstract{
We  analyze a pair of delay-coupled FitzHugh-Nagumo oscillators exhibiting in-out intermittency as a part of the generating mechanism of extreme events. We study in detail the characteristics of in-out intermittency and identify the invariant subsets involved --- a saddle fixed point and a saddle periodic orbit --- neither of which are chaotic as in the previously reported cases of in-out intermittency. Based on the analysis of a periodic attractor possessing in-out dynamics, we can characterize the approach to the invariant synchronization manifold and the spiralling out to the saddle periodic orbit with subsequent ejection from the manifold. Due to the striking similarities, this analysis of in-out dynamics explains also in-out intermittency.
} 
\maketitle
\section{Introduction}
\label{intro}

Intermittency is a well-known phenomenon characterized by an alternation of time periods of regular or laminar dynamics, and irregular or turbulent bursts. The latter are seemingly randomly interspersed in the laminar phase and become more frequent with increasing distance from the transition to intermittency~\cite{Elaskar2017}. Starting from early studies involving fluctuating velocities in turbulent flows~\cite{Batchelor238}, the importance of intermittency has been well-recognized in several physical phenomena including forced nonlinear oscillators~\cite{PhysRevLett.110.184502}, plasma physics~\cite{doi:10.1063/1.1691453,doi:10.1063/1.3385796}, Rayleigh-B\'ernard convection~\cite{PhysRevLett.51.1446,MALASOMA2003487}, electronic circuits~\cite{doi:10.1142/S0218127408021178} and nonlinear Schr\"odinger equations~\cite{doi:10.1063/1.2768513}. Additionally, it has also been used to explain various observed phenomena in economics~\cite{chian2007complex} and medicine~\cite{ZEBROWSKI200474,doi:10.1063/1.4776519}. Early literature~\cite{MANNEVILLE19791,refId0} classifies intermittency into three types, namely I, II and III, based on the Floquet multipliers of the system or the eigenvalues in the local Poincar\'e map~\cite{marek1995chaotic,nayfeh2008applied,rasband2015chaotic,schuster2006deterministic}. Further studies have extended the classification into intermittencies of types V and X, eyelet intermittency and ring intermittency~\cite{PhysRevLett.68.553,PRICE199129,PhysRevLett.81.321,PhysRevLett.97.114101}.

Two classes of intermittency which occur in systems possessing an invariant manifold in their state space are on-off intermittency and in-out intermittency. While the phenomenon of on-off intermittency has been known for a long time and has been well studied~\cite{PhysRevLett.70.279,PhysRevLett.79.47,VENKATARAMANI199666,VENKATARAMANI1995173}; in-out intermittencies were discovered later as a generalisation of on-off intermittency. In-out intermittency, which is the subject of this article was first described in maps~\cite{0951-7715-12-3-009}. Subsequently, the phenomenon was also discovered in deterministic ODE and PDE models~\cite{doi:10.1063/1.1374243} as well as in stochastic systems~\cite{PhysRevE.64.066204}. In fact, it has also been shown that the presence of noise in a system exhibiting in-out intermittency induces coherence resonance~\cite{GAC2007136}. Various studies have revealed that systems exhibiting in-out intermittency show also other interesting phenomena like riddling~\cite{PhysRevLett.83.2926,doi:10.1063/1.166495,doi:10.1063/1.4954022,doi:10.1063/1.5012134}.

Here, we analyze a pair of identical FitzHugh-Nagumo oscillators which are diffusively coupled to each other by two different time delays. Such a system has been recently shown to generate extreme events~\cite{PhysRevE.95.062219}. Interestingly, the study also indicated that in-out intermittency plays a crucial role in the generating mechanism of these extreme events. In this study, we take a closer look at this type of intermittency and identify the two invariant sets involved in it, one of them is responsible for the ``in'' motion while the other one mediates the ``out'' dynamics. To the best of our knowledge, this is the first example of in-out intermittency in a delay-coupled slow-fast system. Keeping in mind the immense impact of extreme events as well as the ubiquity and importance of multi-scale dynamics and delayed systems, the understanding of the mechanism underlying in-out intermittency in such systems is of crucial importance.

The paper is structured in the following way. In the next section, we give a brief overview of the on-off and in-out intermittency and highlight their difference. In Section 3, we describe our model system and the general structure of the phase space. In Section 4, we outline the overall dynamics of the system in various parameter regimes relevant for this study. This  includes the regimes where in-out intermittency is observed. However, it also includes a regime, which although is not related to intermittency, exhibits ``in-out'' dynamics related to a heteroclinic connection. We compare the dynamical regimes in terms of the ``in-out'' dynamics in Section 5. In Section 6, we thoroughly analyze the system in the parameter regime with the heteroclinic connection as its simplified dynamics helps us identify the various components on the invariant manifold which determine the overall dynamics of a typical trajectory. Thereafter, we extend our understanding to the parameter regimes where in-out intermittency is actually observed. In Section 7, we summarize our results and discuss our conclusions.

\section{Intermittencies in systems with invariant manifolds}
\label{sec:intermittencies}

There are two known types of intermittencies specific to systems which involve an invariant manifold embedded in their phase space; namely on-off intermittency~\cite{PhysRevLett.70.279,Pikovsky1984,ashwin1996attractor} and in-out intermittency~\cite{doi:10.1063/1.1374243}. In this section we briefly describe the known general mechanisms of these intermittencies, highlighting the differences between them.

The primary requirements for on-off intermittency is an invariant manifold and trajectories which enter and exit every sufficiently small neighbourhood of that manifold. One of the ways in which this is manifested is when the system has a skew product structure and there exists a strange attractor on the invariant manifold. In such a scenario, the dynamical equations governing the evolution of the trajectories can be such that the attractor behaves transversally attracting in some regions of the manifold and transversally repelling in some other regions. Therefore, a trajectory may get attracted towards the manifold and spend a long time in the neighbourhood of the transversally attracting part of the strange attractor before moving to the transversally repelling part and getting ejected away from the invariant manifold. Note that both getting attracted towards and getting ejected away from the invariant manifold involves the same strange attractor.

\begin{figure}
  \centering
  \includegraphics[width=\textwidth]{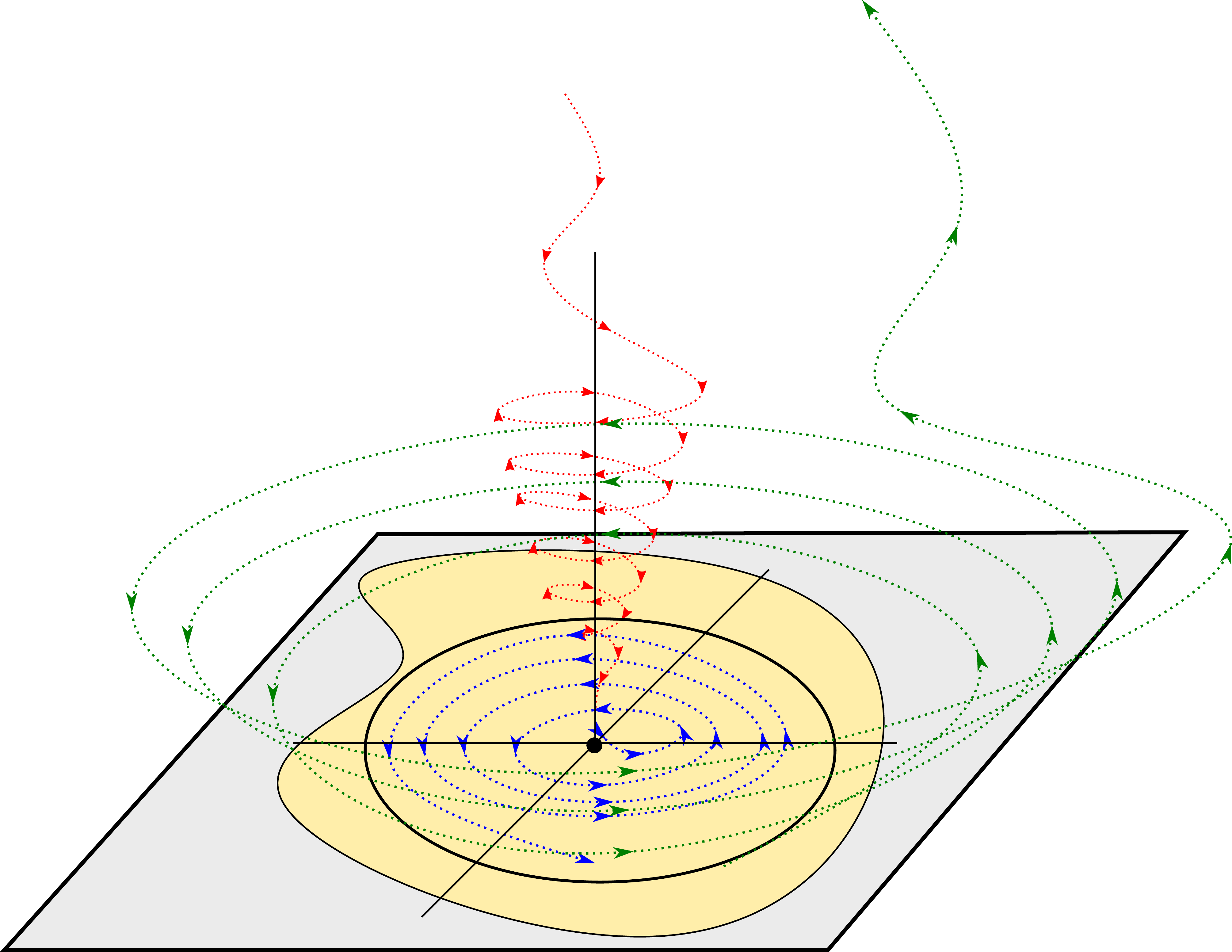}
  \caption{Schematic diagram showing a possible mechanism for in-out intermittency. The gray parallelogram represents the invariant manifold on which, a strange attractor (yellow patch) is present. Embedded in the chaotic attractor is a transversally attracting saddle fixed point (solid black circle) and a transversally repelling saddle limit cycle (black oval curve). The qualitative nature of a typical trajectory is shown by the dotted curves. There are three phases to such a trajectory: From a point not on the invariant manifold, the trajectory (red) gets attracted along the attracting direction of the saddle fixed point. Once, close to the invariant manifold, the trajectory (blue) spirals out of the fixed point along its unstable direction. Since the trajectory is close to the chaotic attractor, its dynamics closely shadows the chaotic attractor until it reaches the transversally repelling unstable limit-cycle and gets ejected (green) away from the invariant manifold from where the mechanism repeats itself until the trajectory converges to one of the many possible attractors of the system.}
  \label{fig:diagram}       
\end{figure}

One of the major differences between on-off and in-out intermittency is that the latter, in general, involves two different invariant sets: one corresponding to attraction towards to the invariant manifold and the other corresponding to repulsion (see Fig.~\ref{fig:diagram}). In other words, the invariant manifold may contain two or even more invariant sets, one of which is transversally stable and the other is transversally unstable. A general trajectory, which is not already on the invariant manifold gets attracted along the transversally stable direction of the first invariant set and stays close to the invariant manifold until it reaches the neighbourhood of the second invariant set which is transversally unstable and is ejected away from the manifold. Note that while strange attractors might be present on the invariant manifold, they do not necessarily play a role either attracting the trajectory towards or repelling it away from the manifold. Note also that, if the system has a skew-product structure, in-out intermittency reduces to on-off intermittency.

In the following section, we introduce a system of two delay-coupled FitzHugh-Nagumo units which shows in-out intermittency. After describing briefly the dynamics of the system, we try to identify the particular invariant manifolds which are involved in generating the intermittency.

\section{Model description}

We analyze the system of delay-coupled FitzHugh-Nagumo (FHN) units which are known to exhibit extreme events~\cite{PhysRevE.95.062219}. It consists of two identical FHN units (indexed by $i=1,2$) coupled to each other using two different diffusive delay couplings. The overall dynamics of the system is given by,
\begin{equation}
  \begin{aligned}
    \dot{x}_i &= x_i(a_i-x_i)(x_i-1)-y_i + \sum_{k=1}^2 M_k (x_j^{(\tau_k)}-x_i) \\
    \dot{y}_i &= b_i x_i - c_i y_i + \sum_{k=1}^2 M_k (y_j^{(\tau_k)}-y_i). 
  \end{aligned}
  \label{eq:Model}
\end{equation}
Here, $x_{i}$ and $y_{i}$ represent the excitatory and inhibitory variables of the $i^{\text{th}}$ FHN unit at any given time $t$, whereas, $x_{i}^{\left( \tau_{k} \right)}$ and $y_{i}^{\left( \tau_{k} \right)}$ represent the corresponding variables at time $t-\tau_{k}$. The internal parameters $a$, $b$ and $c$ are identical for both FHN units and are fixed at $a=-0.025$, $b=0.00652$ and $c=0.02$ for this investigation, where the individual uncoupled FHN units behave as a slow-fast system exhibiting relaxation oscillations. The strength of connection corresponding to time-delay $\tau_{k}$ is given by $M_{k}$. The changing dynamics of the system as the coupling parameters $M_{k}$ and $\tau_{k}$ are varied will be the focus of this paper. Particularly, we are interested in the characteristics of in-out intermittency observed in this system. Since the system contains time-delays, the description of complete dynamics of a trajectory for a given set of parameters is determined by the initial history functions for the dynamical variables $x_{i}$ and $y_{i}$. Here we restrict our analysis to the cases where the initial history functions are constant for $\max\left\{ \tau_{k} \right\} \le t \le 0$. Hence, the initial history function is completely defined by the choice of the initial conditions for $x_{i}$ and $y_{i}$ at time $t=0$.

The identical nature of the two oscillators introduces an exchange symmetry in the system. This symmetry is manifested in the phase space as an invariant manifold corresponding to complete synchrony of the two oscillators. Mathematically, such an invariant synchronization manifold (hereafter referred to as ISM) is given by $x_{1}\left( t-t^{\prime} \right) = x_{2}\left( t-t^{\prime} \right), y_{1}\left( t-t^{\prime} \right) = y_{2}\left( t-t^{\prime} \right)$ for all $t^{\prime} \in \left[ 0,\max\left\{ \tau_{k} \right\} \right]$. In other words, at any given time, the system is said to be on the ISM if the two oscillators have identical current dynamical states and identical histories for up to $\max\left\{ \tau_{k} \right\}$ time in the past. Note that this also implies that the synchronization manifold is infinite dimensional.

\section{Overview of the dynamics}

The system described by Eq.~\ref{eq:Model} can exhibit a variety of different dynamical regimes and can lead to interesting basin structures which have been described in detail in recent literature~\cite{PhysRevE.95.062219,doi:10.1063/1.5012134,saha2017intermittency}. In this article, we investigate the underlying mechanisms of some of those dynamical regimes where the trajectory of the system repeatedly approaches the ISM in the phase space. To this end, in this section, we present the following three dynamical regimes which are classified on the basis of their long term dynamics.

\subsection{Regime 1: Chaotic dynamics exhibiting extreme events}
  
\begin{figure}
  \centering
  \includegraphics[width=\textwidth]{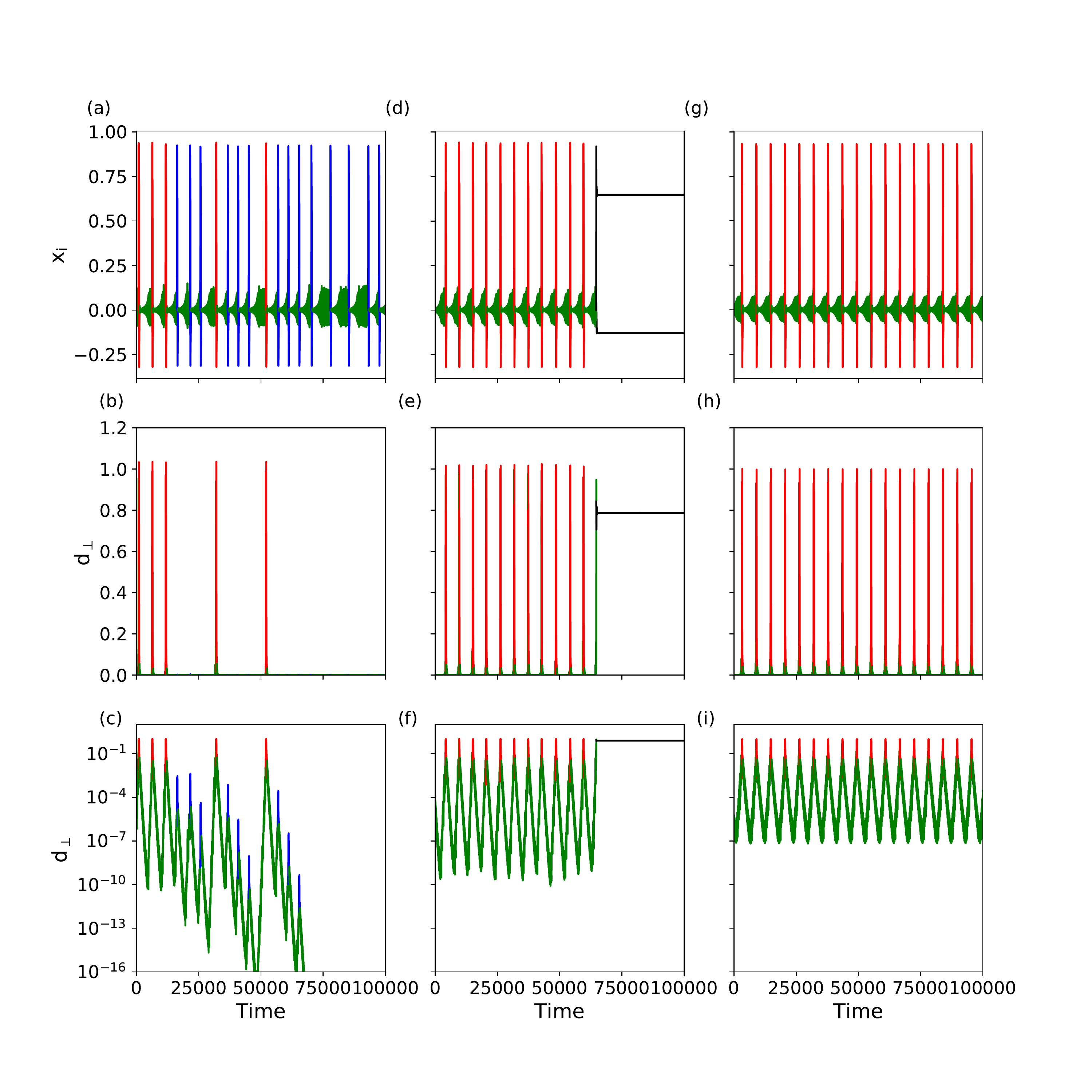}
  \caption{Time-evolution of the dynamical variables $x_{1,2}$ and the distance from the synchronization manifold $d_{\perp}$ for typical trajectories from the three dynamical regimes. (a-c) show above-mentioned properties for a trajectory which the long-term dynamics comprises of chaotic oscillations on the ISM. (d-f) show them for a trajectory which finally converges to one of the stable fixed points outside the ISM. For the trajectory shown in (g-i), the long-term dynamics is a periodic orbit with parts both in the neighborhood of the ISM and also far away from it. Parameters used: $M_{1}=0.005$, $\tau_{1}=80$, $M_{2}=0.0053$, $\tau_{2}=70$ for (a-c); $M_{1}=0.005$, $\tau_{1}=80$, $M_{2}=0.0059$, $\tau_{2}=70$ for (d-f); $M_{1}=0.01$, $\tau_{1}=70$, $M_{2}=0$ for (g-i).  Color code for the trajectories: Small amplitude oscillations in green, synchronous large amplitude oscillations in blue and asynchronous large amplitude oscillations in red.}
  \label{fig:intermittency}       
\end{figure}

\begin{figure}
  \centering
  \includegraphics[width=\textwidth]{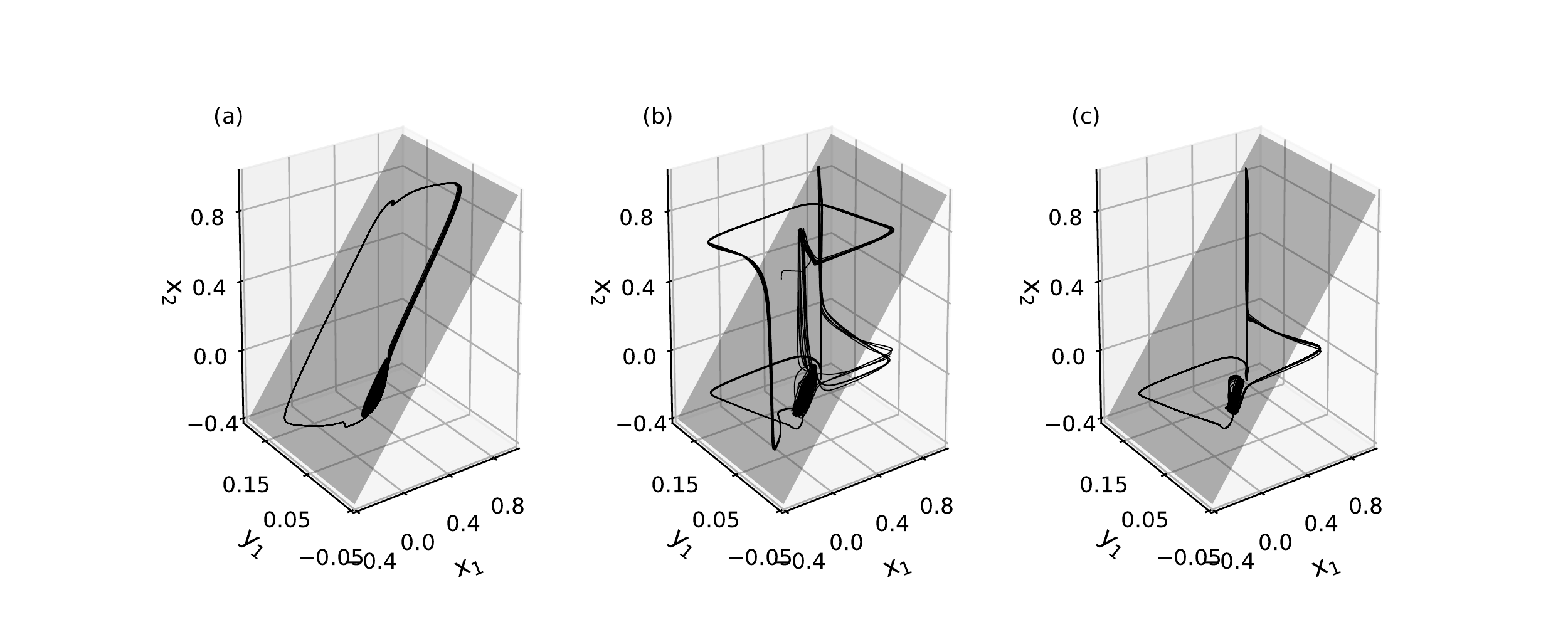}
  \caption{ Three dimensional phase space representations of the trajectories in the three dynamical regimes. The gray surfaces represent the invariant synchronization manifold. (a) and (c) represent the long term dynamics of the trajectories which converge to the chaotic attractor on the ISM (Regime 1) and the periodic orbit (Regime 3) respectively. (b) also shows part of the transients of the trajectory before it converges to the stable fixed point (Regime 2). Parameters used: $M_{1}=0.005$, $\tau_{1}=80$, $M_{2}=0.0053$, $\tau_{2}=70$ for (a); $M_{1}=0.005$, $\tau_{1}=80$, $M_{2}=0.0059$, $\tau_{2}=70$ for (b); $M_{1}=0.01$, $\tau_{1}=70$, $M_{2}=0$ for (c).}
  \label{fig:regimes}       
\end{figure}

If we fix the coupling parameters of the system at $M_{1}=0.005$, $M_{2}=0.0053$, $\tau_{1}=80$ and $\tau_{2}=70$, we have a stable chaotic attractor on the ISM, which comprises of small- and large-amplitude oscillations. Any trajectory which starts on the ISM, converges to the chaotic attractor after an extremely short transient and exhibits extreme events~\cite{PhysRevE.95.062219}. The trajectories which do not start on the ISM, on the other hand, show very long transients before converging to the chaotic attractor (see Fig.~\ref{fig:intermittency}a). Any such trajectory approaches the ISM and thereby executes a few small-amplitude oscillations near that manifold. Thereafter, the trajectory undergoes a single large-amplitude oscillation which can be of one of the two types: (a) \emph{synchronous}, where the trajectory remains close to the ISM during the oscillation and the two oscillators remain in synchrony; or (b) \emph{asynchronous}, where the trajectories are ejected away from the ISM and the two oscillators are out of synchrony during the oscillation. Synchrony between the oscillators is restored after the large amplitude oscillation when the trajectory returns to the neighborhood of the manifold. Due to the chaotic nature of the transient, there seems to be no apparent way to predict if a given large-amplitude oscillation would be synchronous or asynchronous before its occurrence. After the long transient, the trajectory finally converges to the ISM and executes chaotic oscillations corresponding to extreme events manifested as synchronous large-amplitude oscillations (see Fig.\ref{fig:regimes}a).

In order to characterize the repeated approaches towards the ISM during the transient, we define a measure $d_{\perp}$ corresponding to the distance of the trajectory from the manifold as follows,
\begin{equation}
  d_{\perp} = \sqrt{\left( x_{1} - x_{2} \right)^{2} + \left( y_{1} - y_{2} \right)^{2}}.
  \label{eq:dist_synch}
\end{equation}
We can clearly see from Fig.~\ref{fig:intermittency}b,  that $d_{\perp}$ remains close to zero during the small-amplitude and synchronous large-amplitude oscillations. It shows large peaks only during the asynchronous large-amplitude oscillations. This is expected, as the trajectory of the system leaves the neighborhood of the ISM only during asynchronous large-amplitude oscillations. However, analyzing the evolution of $d_{\perp}$ on a semilogarithmic scale (see Fig.~\ref{fig:intermittency}c) gives us further insights into the evolution of $d_{\perp}$ as it switches from near-zero values to large peaks and vice verse. From the figure, it can be seen that during the approach of the trajectory towards the ISM, the amplitude of oscillations of $d_{\perp}$ decreases exponentially. Similarly, during the ejection of the trajectory away from the ISM, the amplitude of oscillations of $d_{\perp}$ increases exponentially. This is very similar to previous results which report the existence of in-out intermittency in other systems~\cite{doi:10.1063/1.1374243,0951-7715-12-3-009,ASHWIN200430,blackbeard2014synchronisation}. Also note that, the trajectories are repelled away from the synchronization manifold for both synchronous and asynchronous large-amplitude oscillations. However, peaks in values of $d_{\perp}$ corresponding to synchronous large-amplitude oscillations are visible only in the semilogarithmic plot because in such a case the trajectory travels only to a small distance away from the manifold before being attracted back towards it.

\subsection{Regime 2: Convergence to steady state}

The second dynamical regime in which we are interested in this paper is observed for $M_{1}=0.005$, $M_{2}=0.0058$, $\tau_{1}=80$ and $\tau_{2}=70$. For this set of parameters, the chaotic attractor on the ISM, which was stable in the previous regime loses its stability. Additionally, a pair of fixed points which lie on either side of the ISM, and were unstable in the previous dynamical regime, gain stability via a reverse Hopf bifurcation. Consequently, any trajectory not starting on the ISM, has no stable attractor on the manifold to which it can converge. Instead, it converges to one of the newly stabilized fixed points (see Fig.~\ref{fig:intermittency}d and Fig.~\ref{fig:regimes}b).

Although the long term dynamics in this regime is different compared to that of the previous regime, both regimes have long chaotic transients for the trajectories which do not start on the ISM. Similar to the previous regime, such a transient trajectory executes small-amplitude oscillations near the ISM. However, in contrast to the previous regime, the large-amplitude oscillation following these small-amplitude oscillations is \emph{always} asynchronous where the trajectory is ejected away from the ISM and the two oscillators loose synchrony. Similar to the previous regime, the synchrony is regained when the trajectory returns to the neighborhood of the ISM after the large-amplitude oscillation to execute small-amplitude oscillations again.

The repeated approach towards the ISM and the subsequent ejection away from it during the transient is similar to the previous regime (see Fig.~\ref{fig:intermittency}e-f) with the amplitude of $d_{\perp}$ oscillations decreasing or increasing exponentially with time and therefore indicates an in-out intermittency in this regime. However, since the large-amplitude oscillations for this regime are always asynchronous, $d_{\perp}$ reaches high values for each large-amplitude oscillation.

Due to the invariant nature of the synchronization manifold, a trajectory that starts on the synchronization manifold cannot leave it. Instead, such a trajectory converges to a small-amplitude limit-cycle on the ISM. The amplitude of oscillations when a trajectory is on such a limit-cycle is similar to that of the small-amplitude oscillations performed by a trajectory which does not start on the ISM. Note that, since this small-amplitude limit-cycle can only be reached by a trajectory starting on the ISM, it can be concluded that the limit-cycle is attracting along the ISM and repelling in the directions transverse to it.

\subsection{Regime 3: Periodic oscillations}

The dynamics in the previous two regimes was characterized by an in-out intermittency followed by convergence to either a chaotic attractor located in the ISM or a stable fixed point outside the manifold. During the intermittency, it was noted that the trajectories repeatedly visit the neighborhood of the ISM in a way that the amplitude of oscillations of $d_{\perp}$ alternatingly increases and decreases exponentially with time. However, such an evolution of $d_{\perp}$ is chaotic and cannot be observed for an arbitrarily long period of time since it is a part of the transient. This, in turn, makes the further analysis of in-out intermittency in this system rather difficult. Therefore, we now present the dynamics of the system in a parameter regime where the alternating exponential increase and decrease of $d_{\perp}$ in time is manifested during the long-term dynamics of the system and hence, can be observed for an arbitrarily long time. Moreover, the long-term time-evolution of $d_{\perp}$ is this regime is periodic, which makes its further analysis much easier.

The above-mentioned properties of $d_{\perp}$ can be observed by fixing the coupling parameters at $M_{1}=0.01$, $\tau_{1}=70$ and $M_{2}=0$. For such a parameter setting, the long-term dynamics of a trajectory which does not start on the ISM is a stable limit-cycle which comprises of small-amplitude oscillations near the ISM and large amplitude oscillations far away from it combining both to a mixed mode oscillation (see Fig.~\ref{fig:intermittency}g and Fig.\ref{fig:regimes}c). As a result, the trajectory visits the neighborhood of the ISM periodically in the long term (see Fig.~\ref{fig:intermittency}h-i). On the other hand, a trajectory starting on the synchronization manifold converges to a small-amplitude limit-cycle on the ISM. Similar to the previous regime, the limit-cycle on the ISM can only be approached by a trajectory starting on the ISM. Hence, the limit-cycle on the ISM is transversally unstable.

Note that, there are two stable limit-cycles in the system which are symmetrically placed on either sides of the ISM. Depending on the initial conditions, a trajectory not starting on the ISM may converge to either of the two limit-cycles. Additionally, any large-amplitude oscillation executed by a trajectory which has converged to either of the two stable limit-cycles, is always asynchronous. We will analyze this simplified form of dynamics in detail in the next section to understand the in-out intermittency observed in the previous regimes.

\section{Analyzing the trajectories}

\begin{figure}
  \centering
  \includegraphics[width=\textwidth]{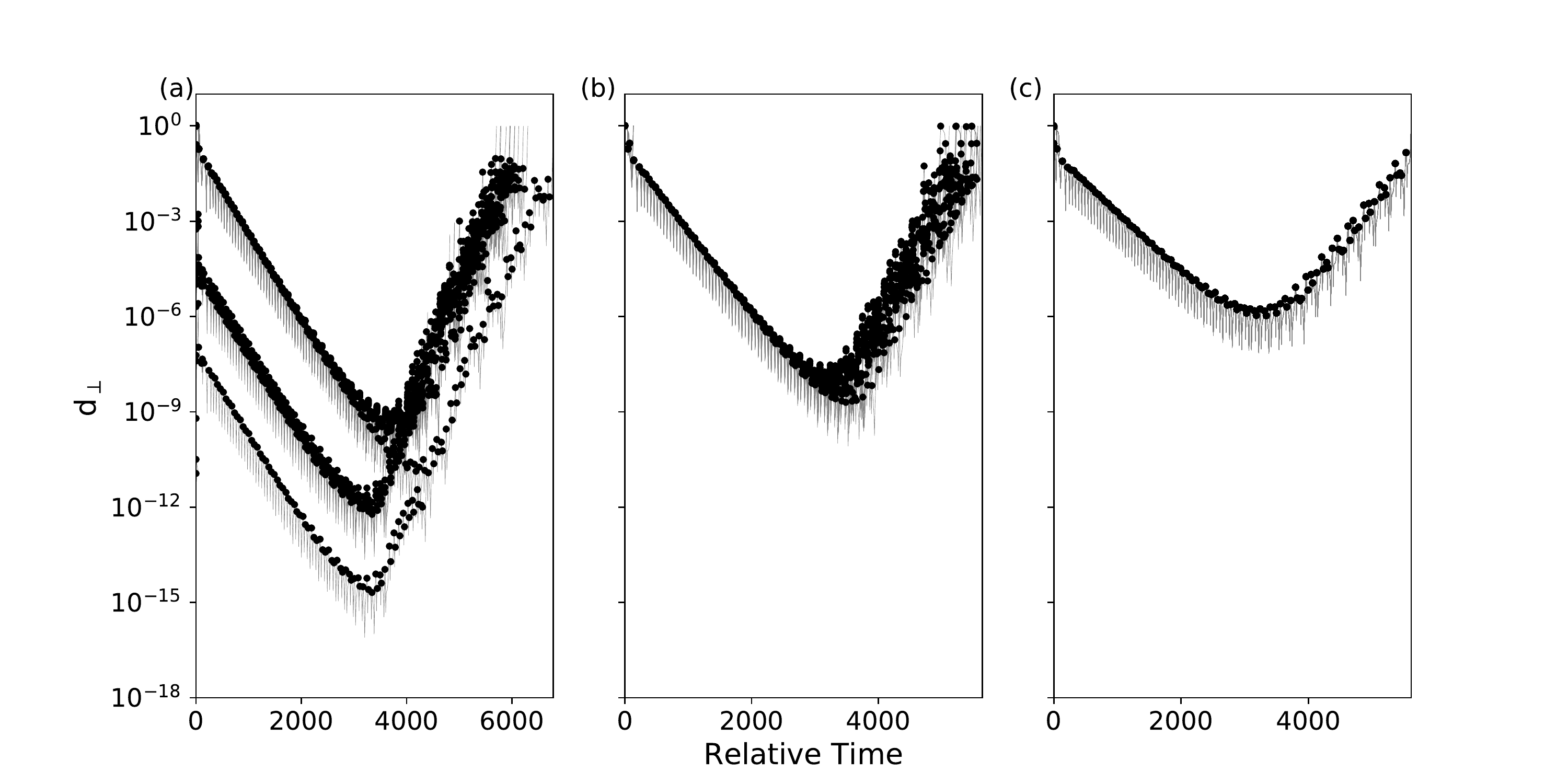}
  \caption{Evolution of a typical trajectory during the excursion cycles superimposed on each other. (a), (b) and (c) show the superimposition for Regimes 1, 2 and 3 respectively. The thin grey lines represent the trajectories and the solid black dots mark the local maximas.}
  \label{fig:overlap}       
\end{figure}

From Fig.~\ref{fig:intermittency} we can see that the transients of the first two dynamical regimes and the long-term dynamics of the third dynamical regime comprises of small- and large-amplitude oscillations. Just after any large-amplitude oscillation, the trajectory is attracted towards the ISM for a certain time. Thereafter, the trajectory is repelled away from the ISM for the remaining time until the next large-amplitude oscillation. The entire trajectory might be seen as being comprised of these excursion cycles which occur between any pair of consecutive large-amplitude oscillations. In order to analyze the mechanism underlying these excursion cycles, we slice the timeseries in Fig.~\ref{fig:intermittency} at each large-amplitude oscillation and superimpose the evolution of $d_{\perp}$ during each excursion cycle on top of each other. The resulting plot for the three regimes is shown in Fig.~\ref{fig:overlap}.

From the figure, it is visible that although the overall rates of increase or decrease of amplitude of $d_{\perp}$ is exponential for all the three regimes, there are notable differences between the excursion cycles of the regimes. For example, since the excursion cycles are part of the chaotic transient in the first two regimes, the trajectory for each excursion cycle is unique. Hence, the different trajectories in Fig.~\ref{fig:overlap}a-b are not completely on top of each other. By contrast, since the excursion cycles are part of the long-term periodic dynamics of the trajectory in the third dynamical regime, all the trajectories in Fig.~\ref{fig:overlap}c completely overlap with each other. Moreover, since the transient in the first regime comprises of both synchronous and asynchronous large-amplitude oscillations, the trajectories may start and end at significantly different values of $d_{\perp}$. This is also reflected in Fig.~\ref{fig:overlap}a, the $d_{\perp}$ values at the beginning or the end of the excursion cycle depend on the type of large-amplitude oscillation preceding or succeeding the cycle.

\section{The mechanism}

\begin{figure}
  \centering
  \includegraphics[width=\textwidth]{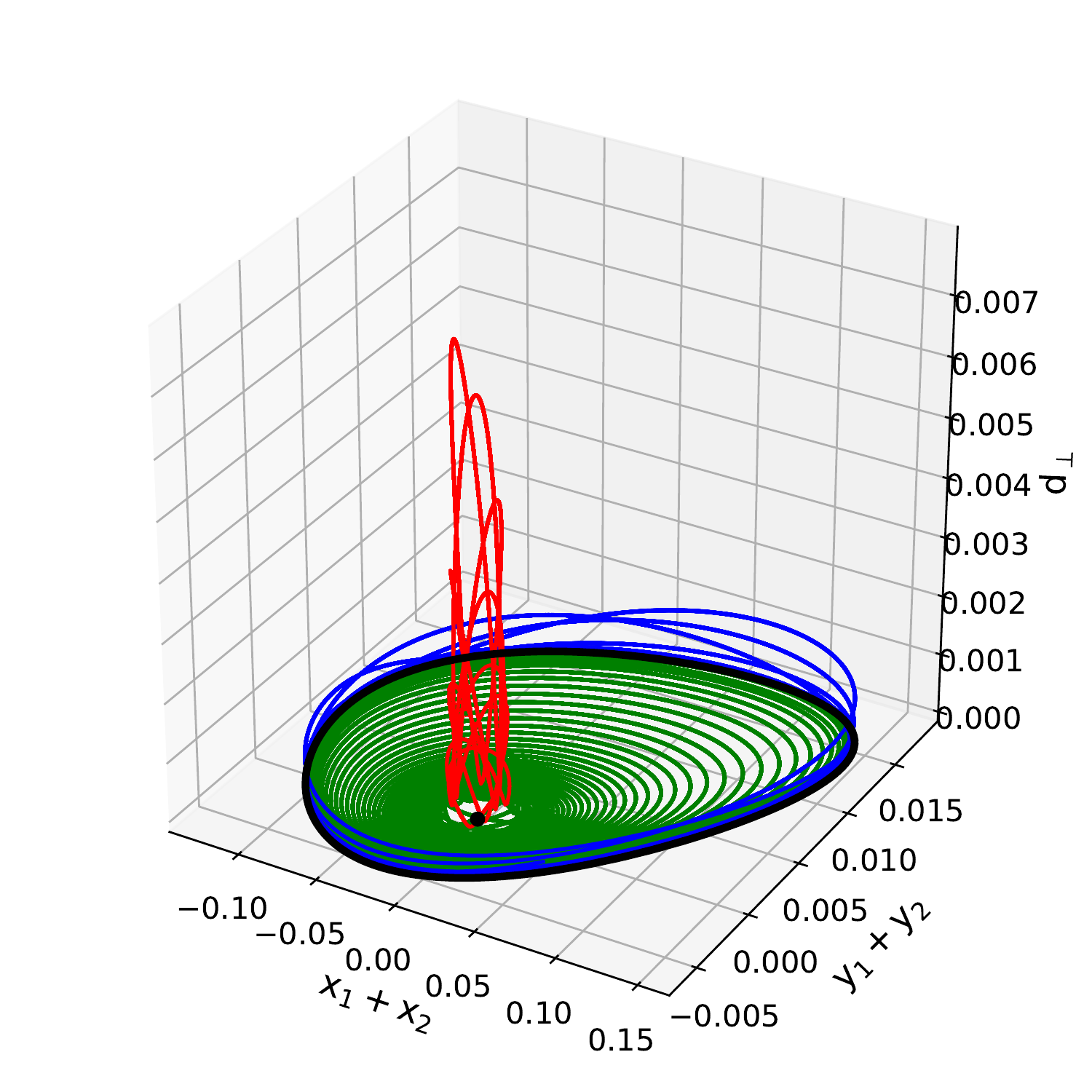}
  \caption{Parts of trajectories in phase space when they are very close to the ISM. Plotted on the $x-$, $y-$ and $z-$ axes are the variables $x_{1}+x_{2}$, $y_{1}+y_{2}$ and the distance metric $d_{\perp}$ respectively. The origin is shown as a filled black circle and the transversally repelling limit-cycle as a black line. The part of the trajectories attracted to the origin are shown in red, while the part where they spiral out from the origin and towards the transversally repelling limit-cycle are shown in green. The final parts of the trajectory as they get repelled from the limit-cycle and the ISM are shown in blue.}
  \label{fig:analysis}       
\end{figure}

\begin{figure}
  \centering
  \includegraphics[width=0.49\textwidth]{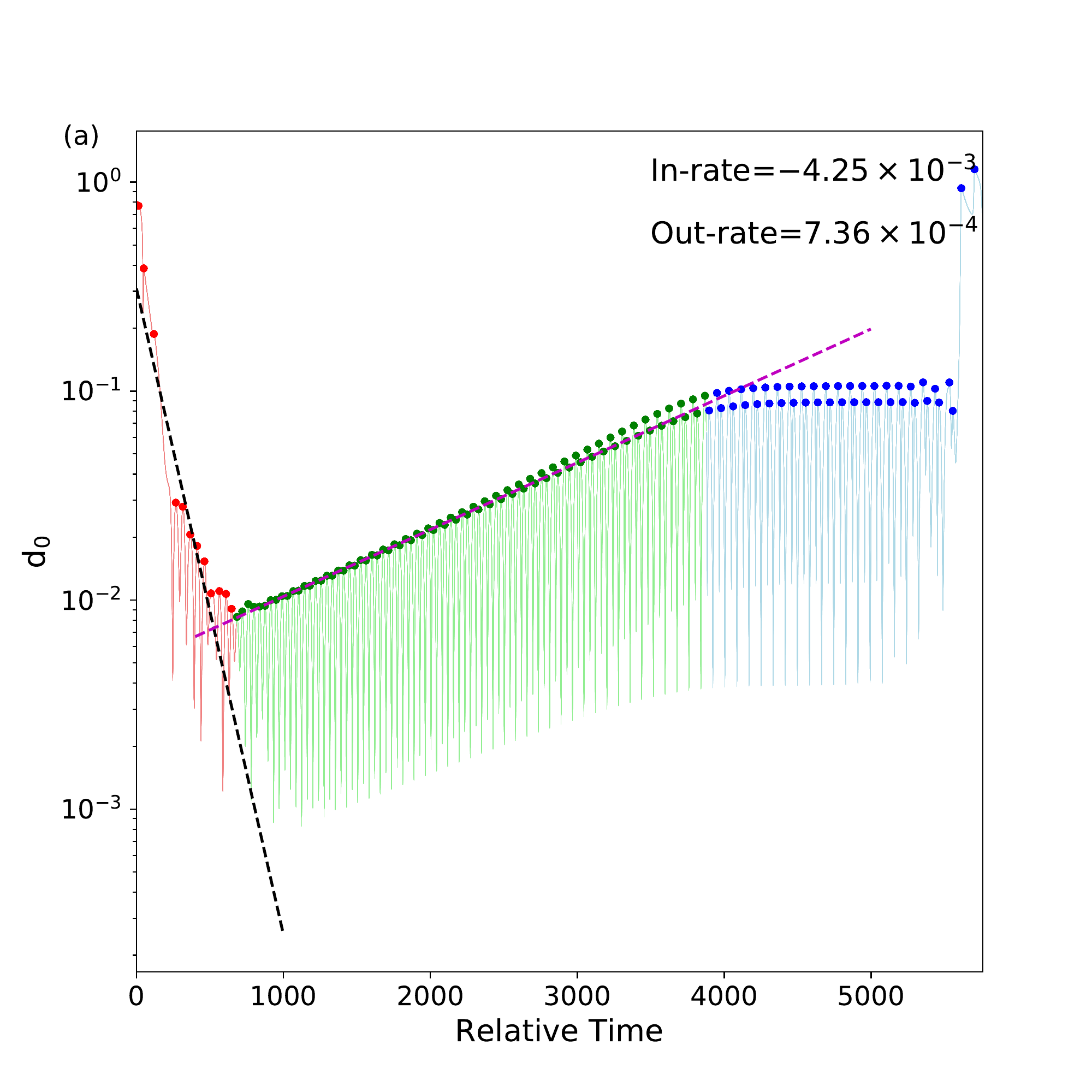}
  \includegraphics[width=0.49\textwidth]{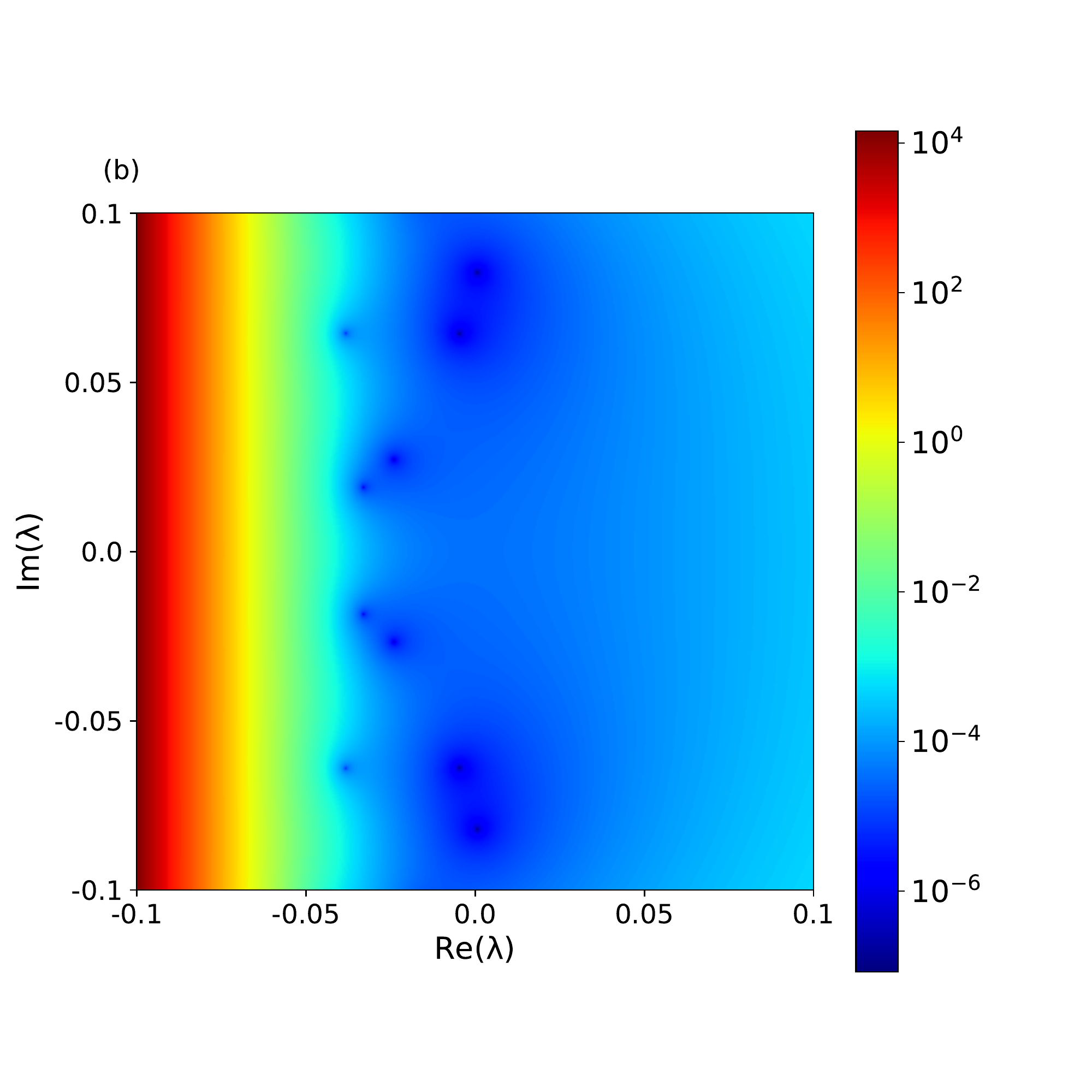}
  \includegraphics[width=0.49\textwidth]{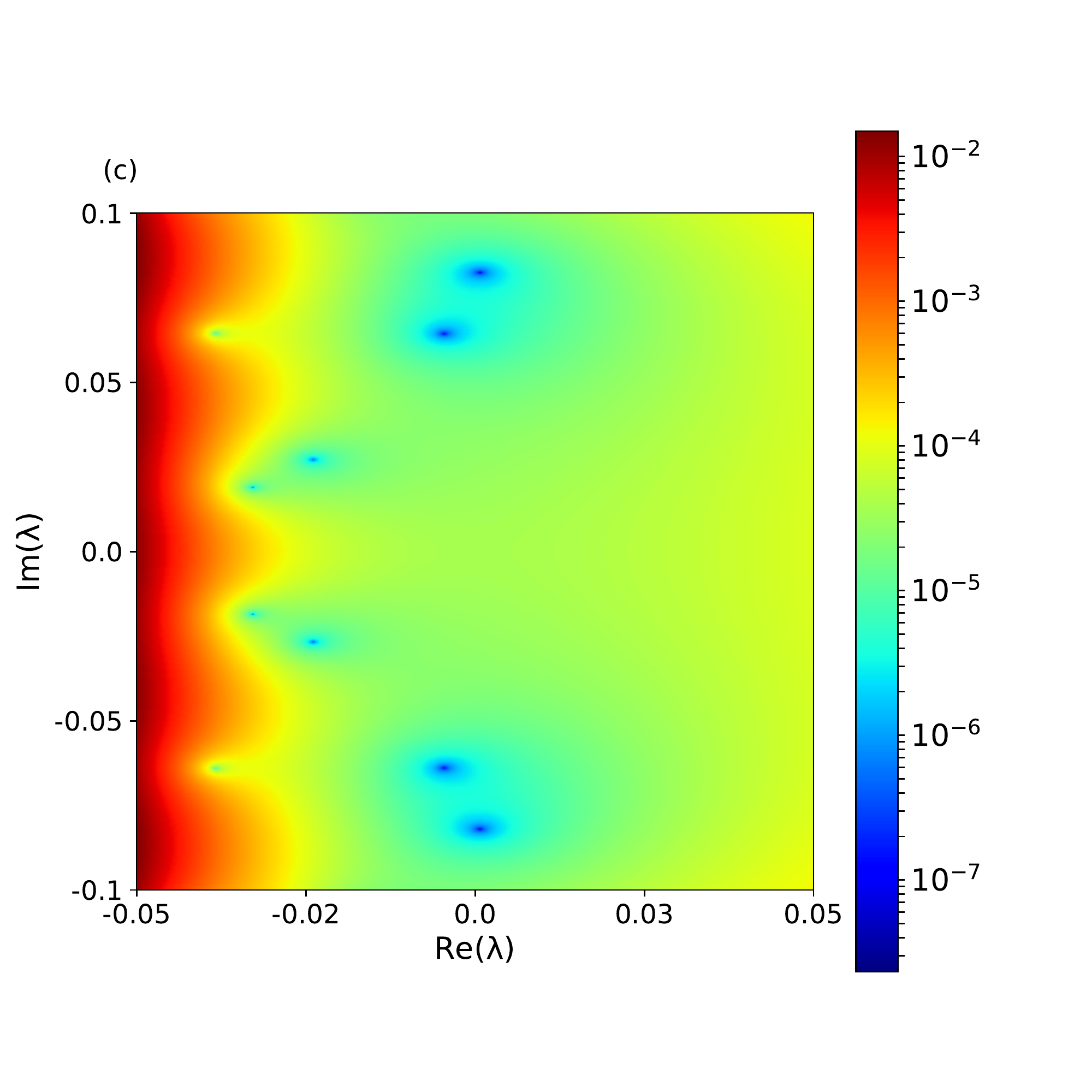}
  \includegraphics[width=0.49\textwidth]{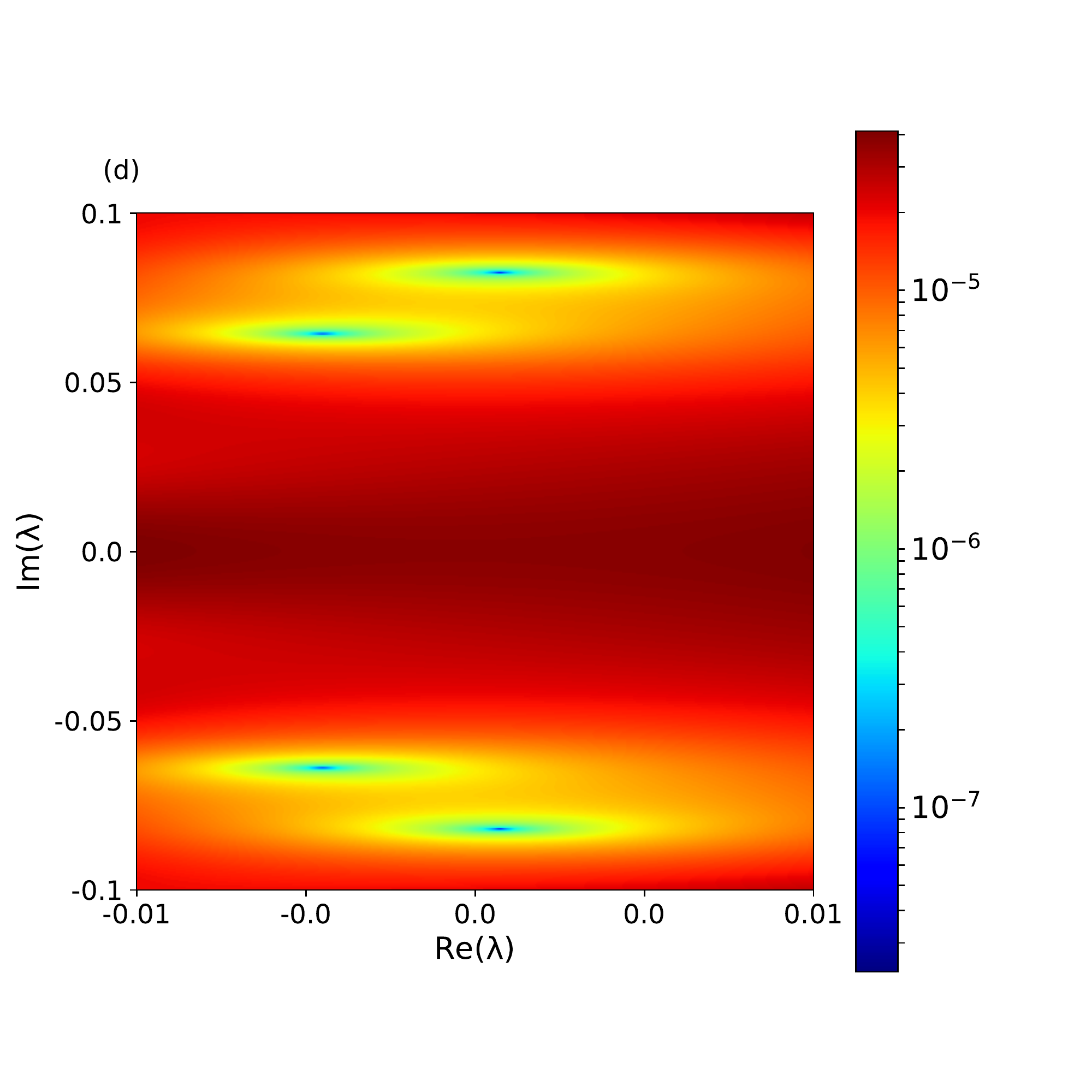}
  \caption{(a) shows the evolution of the distance to the fixed point $d_{0}$ during the excursion cycles. The colors of the various parts are according to the colors in Fig.~\ref{fig:analysis}. The black and magenta dashed lines show the fits of the slopes as the trajectory approaches the origin and spirals out of it respectively. (b) shows the region in the complex plane of $\lambda$ which is scanned for eigenvalues of the origin. Color-coded are the values of the magnitude of the characteristic function $f\left( \lambda \right)$ as defined in Eq.~\ref{eq:characteristic}. (c) and (d) show close-ups of (b).}
  \label{fig:eigenvalues}       
\end{figure}

Despite the differences outlined in the previous section, the evolution of $d_{\perp}$ during the excursion cycles is qualitatively similar for all the regimes presented in this paper. This indicates a similarity of the underlying mechanisms in the three regimes and the invariant sets responsible for them. In this section, we identify these invariant sets for the third regime. We then extrapolate the results obtained to the other two regimes where the excursion cycles are part of the in-out intermittency and the dynamics is more complex.

In Fig.~\ref{fig:analysis} we plot a typical trajectory in the third dynamical regime whenever it visits the neighborhood of the ISM. For each visit, the trajectory undergoes three distinct stages of evolution: (a) While entering the neighborhood of ISM, the trajectory (shown in red) comes very close to the origin which is an unstable fixed point of the system. (b) Thereafter, it spirals out from the origin (shown in green) while staying close to the ISM. (c) As it spirals out, it enters the neighborhood of a small-amplitude limit-cycle on the ISM. Thereafter, instead of spiralling out further, the trajectory is repelled away from the ISM (shown in blue). From numerical simulations, we confirm that the small-amplitude limit-cycle involved here is indeed the same transversally unstable limit-cycle to which any trajectory, starting on the ISM, converges.

From the analysis of trajectories near the ISM, we can conclude that the evolution of the trajectory near the ISM is shaped by the origin which is transversally attracting, but is repelling along the ISM; and a limit-cycle which is attracting along the ISM but is repelling in directions transverse to it. The trajectory comes close to the origin along its stable manifold and then is repelled away from it along its unstable manifold. Subsequently, the trajectory approaches the neighborhood of the limit-cycle and is attracted towards it along its stable manifold before being ejected away from the ISM along its unstable manifold.

To verify the conclusions made from the numerical simulations, we compute the distance of the trajectory from the origin $d_{0}$ (see Fig.~\ref{fig:eigenvalues}a) for each excursion cycle and then superimpose them on each other similar to Fig.~\ref{fig:overlap}. As expected, the trajectories from each excursion cycle completely overlap due to the periodic nature of the dynamics. Moreover, the three stages of evolution of the trajectory as described in the previous paragraphs are clearly seen in the plot. The value of $d_{0}$ decreases sharply as the trajectory approaches the origin, and then increases exponentially as the trajectory spirals out from the origin. After the trajectory reaches the neighborhood of the limit-cycle, it is ejected almost `vertically' away from the ISM. During this vertical ejection, $d_{0}$ remains almost constant over time. This is because, at this stage of evolution, distance of the trajectory from the origin along the ISM is much larger than the distance of the trajectory from the ISM. Therefore, relatively small changes in the distance of the trajectory from the ISM does not affect the distance of the trajectory from the origin.

From Fig.~\ref{fig:eigenvalues}a it can be clearly seen that the rates of evolution of $d_{0}$ are exponential in the neighborhood of the origin. Hence, in order to verify our hypothesis that the origin is involved in pulling the trajectory towards the ISM, the rates of exponential decay and subsequent exponential growth of $d_{0}$ should match the eigenvalues of the fixed point at the origin~\cite{michiels2007stability}. Note that since Eq.~\ref{eq:Model} is a set of delay differential equations, the number of eigenvalues of the fixed point is infinite. However, any eigenvalue $\lambda$ of the origin is the root of the characteristic function,
\begin{equation}
  f(\lambda)=
  \left|
  \begin{matrix}
    -a-M-\lambda & -1 & D & 0\\
    b & -c-M-\lambda & 0 & D\\
    D & 0 & -a-M-\lambda & -1\\
    0 & D & b & -c-M-\lambda
  \end{matrix}
  \right|,
  \label{eq:characteristic}
\end{equation}
where $D = M_{1} e^{-\lambda \tau_{1}} + M_{2} e^{-\lambda \tau_{2}}$ and $M = M_{1} + M_{2}$. Substituting the values of the system parameters in this function and equating it to zero gives us a transcendental equation in $\lambda$. Since the roots of the resulting equation could not be found in closed form, we obtain the roots of the equation by brute force. Since, the roots of the equation could be complex numbers, we perform a parameter scan in a region where $-0.1 \le \Re \left( \lambda \right) \le 0.1$ and $-0.1 \le \Im \left( \lambda \right) \le 0.1$ using a grid of size $10^{6} \times 10^{6}$ and compute the value of $f \left( \lambda \right)$. Thus plotting $\left| f\left( \lambda \right) \right|$ for each grid point (see Fig.~\ref{fig:eigenvalues}(b-d)) and identifying the points where $\left| f\left( \lambda \right) \right| \approx 0$  gives us a close approximation of all the eigenvalues of the origin within the scanning space of $\lambda$.

Adopting the method as described above, we identify 5 pairs of complex conjugate eigenvalues of the origin in the specified scanning space. Two of these pairs which have the largest values of $\Re \left( \lambda \right)$ are $\lambda_{1,2} = 7.29 \times 10^{-4} \pm 0.0822 i$ and $\lambda_{3,4} = -4.50 \times 10^{-3} \pm 0.0642 i$. These values of $\lambda$ are very close to the numerically observed rates of exponential growth and decay of $d_{0}$ when the trajectories are close to the origin. This close correspondence between the numerically observed rates and the eigenvalues of the origin reaffirms the role which the origin plays in our proposed mechanism. Note that a similar analysis can be performed for the small-amplitude limit-cycle, but it would involve the computation of Floquet exponents of the limit-cycle. However, since the computation of Floquet exponents for delay differential equations~\cite{PhysRevE.59.5344} is much more complicated, such an analysis is beyond the scope of this paper.

In the previous paragraphs, we have tried to establish that there are two invariant sets on the ISM which are responsible for the overall dynamics observed in the excursion cycles in dynamical regime 3: (a) the unstable fixed point at the origin responsible for attracting the trajectories towards the ISM and (b) the unstable limit cycle which ejects the trajectories away from the ISM. While, as noted earlier, these excursion cycles cannot be termed intermittency due to their perpetual nature, but the conclusions obtained by this analysis can be extended to the previous two dynamical regimes to understand in-out intermittency.

The unstable fixed point at the origin and the unstable limit-cycle are also present in Regime 2, where the trajectory finally converges to one of the stable fixed points. The presence of these two invariant sets leads to similar excursion cycles. Similar to Regime 3, the transversally repelling limit-cycle responsible for ejection of the trajectories away from the ISM can be reached only by a trajectory which starts on the ISM.

Finally, trajectories in Regime 1 converge to a chaotic attractor on the ISM in the long term. As in the other regimes, the same fixed point at the origin is responsible for trajectories approaching the ISM. However, it is evident that there are many unstable limit-cycles --- at least some of which are transversally repelling --- embedded in the chaotic attractor residing on the ISM. Such transversally repelling limit-cycles are responsible for ejecting the trajectories away from the ISM. However, unlike the other two regimes, these unstable limit-cycles cannot be approached by the trajectories starting on the ISM since any such trajectory converges to the chaotic attractor almost immediately. Therefore, they cannot be directly identified. Further note that, due to the chaotic nature of the attractor on the ISM, a trajectory which does not start on the ISM and undergoes excursion cycles, can spend varying amounts of time close to the chaotic attractor and closely mimicking its dynamics before getting repelled away from the ISM by one of the many transversally repelling limit-cycles. This variation in the lengths of excursion cycles can be seen in Fig.~\ref{fig:overlap}a. Since different unstable limit cycles may push the trajectories in different directions, the trajectories after being repelled by the limit-cycle might go on to execute either a synchronous or an asynchronous large-amplitude oscillation. However, the details of the mechanism distinguishing between the two types of large-amplitude oscillations is beyond the scope of this paper.

\section{Conclusions}

In this study, we present, to the best of our knowledge, the first investigation of in-out intermittency in a system of delay differential equations. We have demonstrated that a system of two identical FHN oscillators connected diffusively to each other using two different delay couplings can exhibit in-out intermittency in at least two distinct parameter regimes.

In both cases, in-out intermittency is a transient phenomenon because the system finally converges to the synchronization manifold or to one of the two fixed points located outside that manifold, respectively. Though the third regime analyzed does not show in-out intermittency, it exhibits a very similar in-out dynamics as part of a limit-cycle attractor. This latter regime allows for a detailed study of the mechanism behind in-out intermittency because of striking similarities with the previous regimes. We identified for all three regimes, the two invariant sets responsible for pulling the trajectory towards the ISM or pushing it away from the ISM; neither of the invariant sets is chaotic. This is in contrast with the previously known examples of in-out intermittency where at least one of the invariant sets involved in the mechanism was a chaotic set. In case of the periodic solution, the observed dynamics is possible because of a heteroclinic connection between the saddle point and the saddle periodic orbit. A remnant of this heteroclinic connection seems to be present also in the other regimes exhibiting in-out intermittency. Note that, chaotic sets are additionally present in the two regimes where in-out intermittency is observed --- as a chaotic attractor in Regime 1 and as a chaotic saddle in Regime 2 --- but they are involved only in making the dynamics chaotic when the trajectory is close to the ISM and not in the mechanism behind the observed in-out intermittency.

As illustrated by Regime 3, we have been able to identify parameter regimes where the essential features of in-out intermittency --- such as the exponential decrease followed by an increase of the distance from the invariant manifold --- can be seen in periodic long-term dynamics. This makes the analysis of the intermittency much easier. In particular, we can estimate the eigenvalues of the saddle fixed point to determine the properties of the approach to and spiralling out of the saddle point. Though these eigenvalues can only be approximated of the periodic regime (Regime 3), the slopes determining the approach to the fixed point and the spiralling out towards the limit-cycle are almost the same for all three regimes, constituting another indicator for the similarity between the in-out periodic dynamics and the in-out intermittency.

Finally, to the best of our knowledge, this is the first demonstration of the phenomenon of in-out intermittency in a system with multiple time scales. In fact, the equations determining the dynamics of the uncoupled FHN units possess two time scales, making each FHN unit a slow-fast system which executes relaxation oscillations. When coupled to each other, the difference between the two time-delays introduces additional time scales in the system. However, the impact of different time scales on the in-out intermittency has not been analyzed in this paper and is left for future research.

\section*{Acknowledgments}

The authors would like to thank P. Ashwin, A. Choudhary and S. Wieczorek for fruitful discussions and critical suggestions. This work was supported by the Volkswagen Foundation (Grant No. 88459).

\end{document}